\documentclass[a4paper,fleqn,usenatbib]{mnras}
\usepackage{newtxtext,newtxmath}

\usepackage[T1]{fontenc}
\usepackage{ae,aecompl}

\usepackage{graphicx}
\usepackage{dcolumn}
\usepackage{bm}
\usepackage{hyperref}
\usepackage{xspace}
\usepackage{soul}
\usepackage[ruled,vlined]{algorithm2e}

\SetAlCapNameFnt{\normalsize}
\SetAlCapFnt{\normalsize}

\usepackage{aas_macros}

\newcommand{\code}{\textsf}

\newcommand{\nlive}{n_\text{live}}
\newcommand{\niter}{n_\text{iter}}
\newcommand{\pvalue}{\text{\textit{p}-value}\xspace}
\newcommand{\pvalues}{\text{\pvalue{}s}\xspace}
\newcommand{\Z}{\mathcal{Z}}
\newcommand{\logZ}{\ensuremath{\log\Z}\xspace}
\newcommand{\like}{\mathcal{L}}
\newcommand{\threshold}{\like^\star}

\newcommand{\intd}{\text{d}}
\newcommand{\params}{\mathbf{\Theta}}
\newcommand{\stoppingtol}{\epsilon}
\newcommand{\efr}{\ensuremath{\code{efr}}\xspace}
\newcommand{\nr}{\ensuremath{n_r}\xspace}

\newcommand{\loggamma}{\ln\Gamma}
\newcommand{\uniform}{\mathcal{U}}
\newcommand{\normal}{\mathcal{N}}

\usepackage{cleveref}

\usepackage[usenames]{xcolor}

\newcommand{\MN}{\code{MultiNest}\xspace}
\newcommand{\PC}{\code{PolyChord}\xspace}
\newcommand{\MNVersion}{\code{\MN-3.12}\xspace}
\newcommand{\PCVersion}{\code{\PC-1.17.1}\xspace}
\newcommand{\anesthetic}{\code{anesthetic}}

\newcommand{\nliveSetting}{1000\xspace}
\newcommand{\tolSetting}{0.01\xspace}
\newcommand{\nrepeatSetting}{100\xspace}
\newcommand{\nrepeatsPerfectSetting}{10,000\xspace}
\newcommand{\nrepeatsMCSetting}{100,000\xspace}
\newcommand{\niterPerfectSetting}{10,000\xspace}

\makeatletter
\def\@printed{}
\def\@journal{}
\def\@oddfoot{}
\def\@evenfoot{}
\makeatother

\title{Nested sampling cross-checks using order statistics}

\author[A. Fowlie et al.]{%
    Andrew Fowlie$^{1}$\thanks{andrew.j.fowlie@njnu.edu.cn},
    Will Handley$^{2,3}$\thanks{wh260@cam.ac.uk},
    and Liangliang Su$^{1}$\thanks{191002001@stu.njnu.edu.cn}
    \\
$^{1}$Department of Physics and Institute of Theoretical Physics, Nanjing Normal University, Nanjing, Jiangsu 210023, China\\
$^{2}$Astrophysics Group, Cavendish Laboratory, J.J.Thomson Avenue, Cambridge, CB3 0HE, UK\\
$^{3}$Kavli Institute for Cosmology, Madingley Road, Cambridge, CB3 0HA, UK
}

\date{}

\pubyear{2020}

\begin{document}
\label{firstpage}
\pagerange{\pageref{firstpage}--\pageref{lastpage}}
\maketitle

\begin{abstract}
Nested sampling (NS) is an invaluable tool in data analysis in modern astrophysics, cosmology, gravitational wave astronomy and particle physics. 
We identify a previously unused property of NS related to order statistics: the insertion indexes of new live points into the existing live points should be uniformly distributed.
This observation enabled us to create a novel cross-check of single NS runs.
The tests can detect when an NS run failed to sample new live points from the constrained prior and plateaus in the likelihood function, which break an assumption of NS and thus leads to unreliable results.
We applied our cross-check to NS runs on toy functions with known analytic results in $2$ -- $50$ dimensions, showing that our approach can detect problematic runs on a variety of likelihoods, settings and dimensions.
As an example of a realistic application, we cross-checked NS runs performed in the context of cosmological model selection.
Since the cross-check is simple, we recommend that it become a mandatory test for every applicable NS run.
\end{abstract}


\section{Introduction}

Nested sampling (NS) was introduced by Skilling in 2004~\citep{2004AIPC..735..395S,Skilling:2006gxv} as a novel algorithm for computing Bayesian evidences and posterior distributions. The algorithm requires few tuning parameters and can cope with traditionally-challenging multimodal and degenerate functions. As a result, popular implementations such as \MN~\citep{Feroz:2007kg,Feroz:2008xx,Feroz:2013hea}, \PC~\citep{Handley:2015fda,Handley:2015xxx} and \code{dynesty}~\citep{2020MNRAS.tmp..280S} have become invaluable tools in modern cosmology~\citep{Mukherjee:2005wg,Easther:2011yq,Martin:2013nzq,Hlozek:2014lca,2013JCAP...02..001A,Akrami:2018odb},
astrophysics~\citep{Trotta:2010mx,2007MNRAS.377L..74L,Buchner:2014nha},
gravitational wave astronomy~\citep{Veitch:2014wba,TheLIGOScientific:2016src,TheLIGOScientific:2016pea,Ashton:2018jfp},
and particle physics~\citep{Trotta:2008bp,Feroz:2008wr,Buchmueller:2013rsa,Workgroup:2017htr}.
Other NS applications include statistical physics~\citep{PhysRevLett.120.250601,PhysRevX.4.031034,doi:10.1021/jp1012973,PhysRevE.89.022302,PhysRevE.96.043311,doi:10.1063/1.4821761},
condensed matter physics~\citep{PhysRevB.93.174108},
and biology~\citep{10.1093/sysbio/syy050,10.1093/bioinformatics/btu675}.

In this work, we propose a cross-check of an important assumption in NS that works on single NS runs. This improves upon previous tests of NS that required toy functions with known analytic properties~\citep{2014arXiv1407.5459B} or multiple runs~\citep{Higson:2018cqj}. The cross-check detects faults in the compression of the parameter space that lead to biased estimates of the evidence. We demonstrate our method on toy functions and previous NS runs used for model selection in cosmology~\citep{Handley:2019tkm}. We anticipate that the cross-check could be applied as broadly as NS itself.

The paper is structured as follows. After recapitulating the relevant aspects of NS in \cref{sec:intro}, we introduce our approach in \cref{sec:test}. We apply our methods to toy functions and a cosmological likelihood in \cref{sec:examples}. We briefly discuss the possibility of using the insertion indexes to debias NS evidence estimates in \cref{sec:debiasing} before concluding in \cref{sec:conclusions}.

\section{NS algorithm}\label{sec:intro}

To establish our notation and explain our cross-check, we briefly summarize the NS algorithm. For more detailed and pedagogical introductions, see e.g., \citep{Skilling:2006gxv,Feroz:2008xx,Handley:2015fda,2020MNRAS.tmp..280S}. NS is primarily an algorithm for computing the Bayesian evidence of a model in light of data. Consider a model with parameters $\params$. The evidence may be written
\begin{equation}\label{eq:Z}
\Z \equiv \int_{\Omega_\params}  \like(\params) \, \pi(\params) \,\intd \params,  
\end{equation}
where $\pi(\params)$ is a prior density for the parameters and $\like(\params)$ is a likelihood function describing the probability of the observed experimental data. The evidence is a critical ingredient in Bayesian model selection in which models are compared by Bayes factors, since Bayes factors are ratios of evidences for two models,
\begin{equation}
B_{10} \equiv \frac{\Z_1}{\Z_0}.
\end{equation}
The Bayes factor $B_{10}$ tells us hows much more we should believe in model $1$ relative to model $0$ in light of experimental data. For an introduction to Bayes factors, see e.g., \citep{Kass:1995loi}.

NS works by casting \cref{eq:Z} as a one-dimensional integral via the volume variable,
\begin{equation}\label{eq:X}
X(\lambda) = \int_{\like(\params) > \lambda}  \pi(\params) \,\intd \params.
\end{equation}
This is the prior volume enclosed within the iso-likelihood contour defined by $\lambda$. The evidence may then be written as
\begin{equation}\label{eq:Z1d}
\Z = \int_0^1 \like(X) \,\intd X,
\end{equation}
where in the overloaded notation $\like(X)$ is the inverse of $X(\lambda)$. 

The remaining challenge is computing the one-dimensional integral in \cref{eq:Z1d}. In NS we begin from $\nlive$ live points drawn from the prior. At each iteration of the NS algorithm, we discard the point with the smallest likelihood, $\threshold$, and sample a replacement drawn from the constrained prior, that is, drawn from $\pi(\params)$ subject to $\like(\params) > \threshold$. By the statistical properties of random samples drawn from the constrained prior, we expect that the volume $X(\threshold)$ compresses by $t$ at each iteration, where
\begin{equation}\label{eq:t}
\langle \log t \rangle = -\frac{1}{\nlive}.
\end{equation}
This enables us to estimate the volume at the $i$-th iteration by $X_i \equiv X(\threshold_i) = e^{-i/\nlive}$ and write the one-dimensional integral using the trapezium rule,
\begin{equation}\label{eq:Z_sum}
\Z \approx \sum_i \threshold_i \, w_i, \qquad w_i = \tfrac12 \left(X_{i - 1} - X_{i + 1}\right).
\end{equation}
The algorithm terminates once an estimate of the maximum remaining evidence, $\Delta \Z$, is less than a specified fraction, $\stoppingtol$, of the total evidence found,
\begin{equation}
\frac{\Delta \Z}{\Z} < \stoppingtol.
\label{eqn:stop}
\end{equation}
The main numerical problem in an implementation of NS is efficiently sampling from the constrained prior.

\subsection{Sampling from the constrained prior}\label{sec:constrained_prior}

Because rejection sampling from the entire prior would be impractically slow as the volume compresses exponentially, implementations of NS typically employ specialised subalgorithms to sample from the constrained prior. When these subalgorithms fail, the evidences may be unreliable. This was considered the most severe drawback of the NS algorithm in \citep{2018arXiv180503924S}.

One such subalgorithm is ellipsoidal sampling~\citep{Mukherjee:2005wg,Feroz:2007kg}, a rejection sampling algorithm in which the live points are bounded by a set of ellipsoids. Potential live points are sampled from the ellipsoids and accepted only if $\like > \threshold$. Ellipsoidal NS is implemented in \MN~\citep{Feroz:2007kg,Feroz:2008xx,Feroz:2013hea}. For this to faithfully sample from the constrained prior, the ellipsoids must completely enclose the iso-likelihood contour defined by $\threshold$. To ensure this is the case, the ellipsoids are expanded by a factor $1 / \efr$, with $\efr = 0.3$ recommended for reliable evidences.

Slice sampling~\citep{neal} is an alternative scheme for sampling from the constrained prior~\citep{aitken,Handley:2015fda}. A chord is drawn from a live point across the entire region enclosed by the iso-likelihood contour and a candidate point is drawn uniformly from along the chord. This is repeated \nr times to reduce correlations between the new point and the original live point. Slice sampling is implemented in \PC~\citep{Handley:2015fda,Handley:2015xxx}. The recommend number of repeats is $\nr = 2d$ for a $d$-dimensional function.

\subsection{Plateaus in the likelihood}\label{sec:plateaus}

Plateaus in the likelihood function, i.e., regions in which $\like(\params) = \text{const.}$, were discussed in \citep{2004AIPC..735..395S,Skilling:2006gxv} and more recently in \citep{2020arXiv200508602S}. In \citep{2020arXiv200508602S} it was stressed that they can lead to faulty estimates of the compression. In such cases, the live points are not uniformly distributed in $X$ (\cref{eq:X}), violating assumptions in \cref{eq:t}.

\section{Using insertion indexes}\label{sec:test}

\begingroup
\begin{table*}
    \centerline{%
\begin{tabular}{cccccccccc}
$\efr$ & $d$ & Analytic \logZ & Mean $\logZ\pm \Delta\logZ$ & $\sigma_{\logZ}$ & SEM \logZ & Inaccuracy & Bias & Median \pvalue & Median rolling \pvalue\\
\hline
\hyperref[sec:gaussian]{Gaussian}\\
\hline
$0.10$ & $2$ & $0$ & $-0.00\pm0.10$ & $0.10$ & $0.01$ & \textcolor{black}{$-0.04$} & \textcolor{black}{$-0.47$} & \textcolor{black}{$0.50$} & \textcolor{black}{$0.49$}\\
$0.10$ & $10$ & $0$ & $0.01\pm0.23$ & $0.21$ & $0.02$ & \textcolor{black}{$0.04$} & \textcolor{black}{$0.48$} & \textcolor{black}{$0.59$} & \textcolor{black}{$0.60$}\\
$0.10$ & $30$ & $0$ & $0.38\pm0.41$ & $0.36$ & $0.04$ & \textcolor{black}{$0.93$} & \textcolor{red}{$10.56$} & \textcolor{black}{$0.52$} & \textcolor{red}{$2.7\cdot10^{-4}$}\\
$0.10$ & $50$ & $0$ & $2.08\pm0.52$ & $0.50$ & $0.05$ & \textcolor{red}{$3.98$} & \textcolor{red}{$41.25$} & \textcolor{black}{$0.38$} & \textcolor{red}{$4.5\cdot10^{-24}$}\\
$1$ & $2$ & $0$ & $-0.00\pm0.10$ & $0.10$ & $0.01$ & \textcolor{black}{$-0.04$} & \textcolor{black}{$-0.46$} & \textcolor{black}{$0.52$} & \textcolor{black}{$0.49$}\\
$1$ & $10$ & $0$ & $0.57\pm0.23$ & $0.22$ & $0.02$ & \textcolor{black}{$2.43$} & \textcolor{red}{$26.07$} & \textcolor{black}{$0.21$} & \textcolor{red}{$1.2\cdot10^{-4}$}\\
$1$ & $30$ & $0$ & $2.35\pm0.40$ & $0.37$ & $0.04$ & \textcolor{red}{$5.83$} & \textcolor{red}{$63.82$} & \textcolor{black}{$0.23$} & \textcolor{red}{$2.2\cdot10^{-23}$}\\
$1$ & $50$ & $0$ & $4.06\pm0.52$ & $0.44$ & $0.04$ & \textcolor{red}{$7.81$} & \textcolor{red}{$92.99$} & \textcolor{black}{$0.30$} & \textcolor{red}{$1.3\cdot10^{-34}$}\\
$10$ & $2$ & $0$ & $-64.75\pm0.11$ & $93.15$ & $9.31$ & \textcolor{red}{$-532.44$} & \textcolor{red}{$-6.95$} & \textcolor{red}{$7.7\cdot10^{-3}$} & \textcolor{black}{$0.06$}\\
$10$ & $10$ & $0$ & $2.81\pm0.23$ & $0.19$ & $0.02$ & \textcolor{red}{$12.30$} & \textcolor{red}{$150.55$} & \textcolor{red}{$2.1\cdot10^{-6}$} & \textcolor{red}{$1.7\cdot10^{-19}$}\\
$10$ & $30$ & $0$ & $4.30\pm0.40$ & $0.25$ & $0.02$ & \textcolor{red}{$10.75$} & \textcolor{red}{$174.47$} & \textcolor{black}{$0.02$} & \textcolor{red}{$3.1\cdot10^{-68}$}\\
$10$ & $50$ & $0$ & $6.04\pm0.52$ & $0.31$ & $0.03$ & \textcolor{red}{$11.66$} & \textcolor{red}{$197.79$} & \textcolor{black}{$0.08$} & \textcolor{red}{$1.1\cdot10^{-93}$}\\
\hline
\hyperref[sec:rosenbrock]{Rosenbrock}\\
\hline
$0.10$ & $2$ & $-5.80$ & $-5.79\pm0.07$ & $0.07$ & $0.01$ & \textcolor{black}{$0.22$} & \textcolor{black}{$2.32$} & \textcolor{black}{$0.50$} & \textcolor{black}{$0.54$}\\
$1$ & $2$ & $-5.80$ & $-5.72\pm0.07$ & $0.05$ & $0.01$ & \textcolor{black}{$1.18$} & \textcolor{red}{$15.31$} & \textcolor{black}{$0.08$} & \textcolor{black}{$0.32$}\\
$10$ & $2$ & $-5.80$ & $-5.77\pm0.07$ & $0.34$ & $0.03$ & \textcolor{black}{$0.65$} & \textcolor{black}{$1.09$} & \textcolor{red}{$9.6\cdot10^{-3}$} & \textcolor{black}{$0.07$}\\
\hline
\hyperref[sec:shells]{Shells}\\
\hline
$0.10$ & $2$ & $-1.75$ & $-1.75\pm0.05$ & $0.05$ & $0.01$ & \textcolor{black}{$-0.06$} & \textcolor{black}{$-0.64$} & \textcolor{black}{$0.55$} & \textcolor{black}{$0.55$}\\
$0.10$ & $10$ & $-14.59$ & $-14.59\pm0.12$ & $0.13$ & $0.01$ & \textcolor{black}{$0.02$} & \textcolor{black}{$0.16$} & \textcolor{black}{$0.57$} & \textcolor{black}{$0.56$}\\
$0.10$ & $30$ & $-60.13$ & $-59.61\pm0.24$ & $0.21$ & $0.02$ & \textcolor{black}{$2.11$} & \textcolor{red}{$24.29$} & \textcolor{black}{$0.37$} & \textcolor{red}{$7.3\cdot10^{-6}$}\\
$0.10$ & $50$ & $-112.42$ & $-110.15\pm0.33$ & $0.20$ & $0.02$ & \textcolor{red}{$6.87$} & \textcolor{red}{$115.58$} & \textcolor{black}{$0.07$} & \textcolor{red}{$3.7\cdot10^{-23}$}\\
$1$ & $2$ & $-1.75$ & $-1.71\pm0.05$ & $0.05$ & $0.00$ & \textcolor{black}{$0.79$} & \textcolor{red}{$8.52$} & \textcolor{red}{$4.7\cdot10^{-3}$} & \textcolor{black}{$0.10$}\\
$1$ & $10$ & $-14.59$ & $-13.92\pm0.12$ & $0.10$ & $0.01$ & \textcolor{red}{$5.57$} & \textcolor{red}{$65.88$} & \textcolor{black}{$0.02$} & \textcolor{red}{$1.1\cdot10^{-5}$}\\
$1$ & $30$ & $-60.13$ & $-57.57\pm0.24$ & $0.17$ & $0.02$ & \textcolor{red}{$10.67$} & \textcolor{red}{$151.79$} & \textcolor{red}{$7.7\cdot10^{-3}$} & \textcolor{red}{$1.4\cdot10^{-20}$}\\
$1$ & $50$ & $-112.42$ & $-107.97\pm0.33$ & $0.18$ & $0.02$ & \textcolor{red}{$13.63$} & \textcolor{red}{$218.07$} & \textcolor{red}{$3.5\cdot10^{-3}$} & \textcolor{red}{$3.6\cdot10^{-37}$}\\
$10$ & $2$ & $-1.75$ & $-1.73\pm0.05$ & $0.12$ & $0.01$ & \textcolor{black}{$0.39$} & \textcolor{black}{$1.45$} & \textcolor{black}{$0.07$} & \textcolor{black}{$0.18$}\\
$10$ & $10$ & $-14.59$ & $-11.73\pm0.11$ & $0.09$ & $0.01$ & \textcolor{red}{$25.56$} & \textcolor{red}{$321.53$} & \textcolor{red}{$6.8\cdot10^{-18}$} & \textcolor{red}{$1.7\cdot10^{-19}$}\\
$10$ & $30$ & $-60.13$ & $-55.41\pm0.24$ & $0.13$ & $0.01$ & \textcolor{red}{$20.03$} & \textcolor{red}{$367.16$} & \textcolor{red}{$3.0\cdot10^{-6}$} & \textcolor{red}{$9.3\cdot10^{-66}$}\\
$10$ & $50$ & $-112.42$ & $-105.82\pm0.32$ & $0.13$ & $0.01$ & \textcolor{red}{$20.42$} & \textcolor{red}{$480.50$} & \textcolor{red}{$9.3\cdot10^{-6}$} & \textcolor{red}{$2.2\cdot10^{-92}$}\\
\hline
\hyperref[sec:gaussian-log-gamma]{Mixture}\\
\hline
$0.10$ & $2$ & $-8.19$ & $-8.18\pm0.06$ & $0.06$ & $0.01$ & \textcolor{black}{$0.08$} & \textcolor{black}{$0.68$} & \textcolor{black}{$0.47$} & \textcolor{black}{$0.58$}\\
$0.10$ & $10$ & $-40.94$ & $-40.56\pm0.16$ & $0.10$ & $0.01$ & \textcolor{black}{$2.46$} & \textcolor{red}{$38.50$} & \textcolor{black}{$0.18$} & \textcolor{black}{$0.46$}\\
$0.10$ & $20$ & $-81.89$ & $-79.04\pm0.22$ & $0.14$ & $0.01$ & \textcolor{red}{$13.05$} & \textcolor{red}{$210.43$} & \textcolor{red}{$2.7\cdot10^{-6}$} & \textcolor{red}{$1.5\cdot10^{-3}$}\\
$1$ & $2$ & $-8.19$ & $-8.16\pm0.06$ & $0.06$ & $0.01$ & \textcolor{black}{$0.55$} & \textcolor{red}{$5.65$} & \textcolor{black}{$0.34$} & \textcolor{black}{$0.58$}\\
$1$ & $10$ & $-40.94$ & $-38.66\pm0.15$ & $0.08$ & $0.01$ & \textcolor{red}{$15.30$} & \textcolor{red}{$304.65$} & \textcolor{red}{$1.0\cdot10^{-9}$} & \textcolor{red}{$1.7\cdot10^{-4}$}\\
$1$ & $20$ & $-81.89$ & $-76.83\pm0.21$ & $0.13$ & $0.01$ & \textcolor{red}{$23.69$} & \textcolor{red}{$388.70$} & \textcolor{red}{$1.4\cdot10^{-11}$} & \textcolor{red}{$5.6\cdot10^{-7}$}\\
$10$ & $2$ & $-8.19$ & $-8.19\pm0.06$ & $0.19$ & $0.02$ & \textcolor{black}{$0.08$} & \textcolor{black}{$0.15$} & \textcolor{black}{$0.08$} & \textcolor{black}{$0.18$}\\
$10$ & $10$ & $-40.94$ & $-36.74\pm0.14$ & $0.07$ & $0.01$ & \textcolor{red}{$29.36$} & \textcolor{red}{$598.27$} & \textcolor{red}{$6.1\cdot10^{-24}$} & \textcolor{red}{$1.1\cdot10^{-8}$}\\
$10$ & $20$ & $-81.89$ & $-74.70\pm0.21$ & $0.12$ & $0.01$ & \textcolor{red}{$34.51$} & \textcolor{red}{$602.35$} & \textcolor{red}{$4.9\cdot10^{-17}$} & \textcolor{red}{$7.1\cdot10^{-13}$}\\
\end{tabular}
}
\caption{\label{tab:MN_summary} Summary of results of our insertion index cross-check for \MN. The numerical results are the average from \nrepeatSetting runs. Biases and inaccuracies greater than $3$ and \pvalues less than $0.01$ are highlighted by red.}
\end{table*}
\endgroup

By \emph{insertion index}, we mean the index at which an element must be inserted to maintain order in an sorted list. With a left-sided convention, the insertion index $i$ of a sample $y$ in an sorted list $o$ is such that
\begin{equation}\label{eq:insertion_index}
o_{i - 1} < y \le o_i.
\end{equation}
The key idea in this paper is to use the insertion indexes of new live points relative to existing live points sorted by enclosed prior volume, $X$, to detect problems in sampling from the constrained prior. 
Since the relationship between volume and likelihood is monotonic, we can sort by volume by sorting by likelihood. If new live points are genuinely sampled from the constrained prior leading to a uniform distribution in $X$, the insertion indexes, $i$, should be discrete uniformly distributed from $0$ to $\nlive - 1$,
\begin{equation}\label{EQ:UNIFORM}
    i \sim \uniform(0, \nlive - 1).
\end{equation}
This result from order statistics is proven in \cref{app:proof}. During a NS run of $\niter$ iterations we thus find $\niter$ insertion indexes that should be uniformly distributed. Imagine, however, that during a NS run using ellipsoidal sampling, the ellipsoids encroached on the true iso-likelihood contour. In that case, the insertion indexes near the lowest-likelihood live points could be disfavoured, and the distribution of insertion indexes would deviate from uniformity. Alternatively, imagine that the likelihood function contains a plateau. Any initial live points that lie in the plateau share the same insertion index, leading to many repeated indexes and a strong deviation from a uniform distribution.

Thus, we can perform a statistical test on the insertion indexes to detect deviations from a uniform distribution. The choice of test isn't important to our general idea of using information in the insertion indexes, though in our examples we use a Kolmogorov-Smirnov (KS) test~\citep{smirnov1948,kolmogorov1933sulla}, which we found to be powerful, to compute a \pvalue from all the iterations. We describe the KS test in \cref{app:ks}.

Excepting plateaus, deviations from uniformity are caused by a \emph{change} in the distribution of new live points with respect to the existing live points. Since there is no technical challenge in sampling the initial live points from the prior, failures should typically occur during a run and thus be accompanied by a change in the distribution. In runs with many iterations in which a change occurs only once, the power of the test may be diluted by the many iterations before and after the distribution changes, as the insertion indexes before and after the change should be uniformly distributed. To mitigate this, we also perform multiple tests on chunks of iterations, find the smallest resulting \pvalue and apply a correction for multiple testing. We later refer to this as the rolling \pvalue. Since the volume compresses by $e$ in $\nlive$ iterations, we pick $\nlive$ as a reasonable size for a chunk of iterations. We treat each chunk as independent. The procedure for computing the rolling \pvalue is detailed in \cref{algo:rolling_p_value}. For clarity, let us stress that we later present \pvalues from all the iterations and rolling \pvalues. Functionality to perform these tests on \MN and \PC output is now included in \code{anesthetic-1.3.6 }~\citep{Handley:2019mfs}.


\begin{algorithm}[h]
\SetAlgoLined
\KwIn{Set of $\niter$ insertion indexes}
 Split the insertion indexes into consecutive chunks of size $\nlive$. The size of the final chunk may be less than $\nlive$\;
 \ForEach{chunk of insertion indexes}{Apply KS test to obtain a \pvalue\;}
 Let $p$ equal the minimum of such \pvalues\;
 Let $n$ equal the number of chunks\;
 \KwRet{Rolling \pvalue{} --- minimum \pvalue adjusted for multiple tests,~$1 - (1 - p)^n$\;}
 \caption{The rolling \pvalue.}
 \label{algo:rolling_p_value}
\end{algorithm}

We furthermore neglect correlations between the insertion indexes. 
Finally, we stress that the magnitude of the deviation from uniform, as well as the \pvalue, should be noted. A small \pvalue alone isn't necessarily cause for concern, if the departure from uniformity is negligible. 

\section{Examples}\label{sec:examples}

\begingroup
\begin{table*}
    \centerline{%
\begin{tabular}{cccccccccc}
$d/\nr$ & $d$ & Analytic \logZ & Mean $\logZ\pm \Delta\logZ$ & $\sigma_{\logZ}$ & SEM \logZ & Inaccuracy & Bias & Median \pvalue & Median rolling \pvalue\\
\hline
\hyperref[sec:gaussian]{Gaussian}\\
\hline
$0.50$ & $2$ & $0$ & $0.01\pm0.11$ & $0.11$ & $0.01$ & \textcolor{black}{$0.11$} & \textcolor{black}{$1.03$} & \textcolor{black}{$0.54$} & \textcolor{black}{$0.60$}\\
$0.50$ & $10$ & $0$ & $-0.00\pm0.23$ & $0.23$ & $0.02$ & \textcolor{black}{$-0.01$} & \textcolor{black}{$-0.10$} & \textcolor{black}{$0.48$} & \textcolor{black}{$0.52$}\\
$0.50$ & $30$ & $0$ & $-0.06\pm0.41$ & $0.37$ & $0.04$ & \textcolor{black}{$-0.15$} & \textcolor{black}{$-1.61$} & \textcolor{black}{$0.54$} & \textcolor{black}{$0.57$}\\
$0.50$ & $50$ & $0$ & $-0.05\pm0.52$ & $0.59$ & $0.06$ & \textcolor{black}{$-0.10$} & \textcolor{black}{$-0.85$} & \textcolor{black}{$0.58$} & \textcolor{black}{$0.51$}\\
$1$ & $2$ & $0$ & $-0.02\pm0.11$ & $0.10$ & $0.01$ & \textcolor{black}{$-0.19$} & \textcolor{black}{$-1.96$} & \textcolor{black}{$0.42$} & \textcolor{black}{$0.48$}\\
$1$ & $10$ & $0$ & $-0.04\pm0.23$ & $0.18$ & $0.02$ & \textcolor{black}{$-0.17$} & \textcolor{black}{$-2.20$} & \textcolor{black}{$0.55$} & \textcolor{black}{$0.59$}\\
$1$ & $30$ & $0$ & $-0.83\pm0.41$ & $0.40$ & $0.04$ & \textcolor{black}{$-2.06$} & \textcolor{red}{$-20.73$} & \textcolor{black}{$0.61$} & \textcolor{black}{$0.46$}\\
$1$ & $50$ & $0$ & $-2.48\pm0.52$ & $0.46$ & $0.05$ & \textcolor{red}{$-4.73$} & \textcolor{red}{$-54.22$} & \textcolor{black}{$0.49$} & \textcolor{black}{$0.59$}\\
$2$ & $2$ & $0$ & $-0.01\pm0.11$ & $0.15$ & $0.02$ & \textcolor{black}{$-0.12$} & \textcolor{black}{$-0.89$} & \textcolor{black}{$0.47$} & \textcolor{black}{$0.53$}\\
$10$ & $10$ & $0$ & $2.20\pm0.23$ & $0.73$ & $0.07$ & \textcolor{red}{$9.50$} & \textcolor{red}{$30.29$} & \textcolor{black}{$0.13$} & \textcolor{black}{$0.22$}\\
$30$ & $30$ & $0$ & $48.37\pm0.64$ & $6.85$ & $0.69$ & \textcolor{red}{$112.25$} & \textcolor{red}{$70.58$} & \textcolor{red}{$8.2\cdot10^{-10}$} & \textcolor{black}{$0.02$}\\
$50$ & $50$ & $0$ & $69.74\pm3.05$ & $6.55$ & $0.65$ & \textcolor{red}{$23.31$} & \textcolor{red}{$106.51$} & \textcolor{red}{$8.0\cdot10^{-86}$} & \textcolor{red}{$1.4\cdot10^{-6}$}\\
\hline
\hyperref[sec:rosenbrock]{Rosenbrock}\\
\hline
$0.50$ & $2$ & $-5.80$ & $-5.79\pm0.07$ & $0.08$ & $0.01$ & \textcolor{black}{$0.20$} & \textcolor{black}{$1.71$} & \textcolor{black}{$0.42$} & \textcolor{black}{$0.46$}\\
$1$ & $2$ & $-5.80$ & $-5.81\pm0.07$ & $0.08$ & $0.01$ & \textcolor{black}{$-0.05$} & \textcolor{black}{$-0.55$} & \textcolor{black}{$0.44$} & \textcolor{black}{$0.52$}\\
$2$ & $2$ & $-5.80$ & $-5.83\pm0.07$ & $0.09$ & $0.01$ & \textcolor{black}{$-0.36$} & \textcolor{black}{$-2.82$} & \textcolor{black}{$0.56$} & \textcolor{black}{$0.49$}\\
\hline
\hyperref[sec:shells]{Shells}\\
\hline
$0.50$ & $2$ & $-1.75$ & $-1.74\pm0.05$ & $0.05$ & $0.00$ & \textcolor{black}{$0.16$} & \textcolor{black}{$1.54$} & \textcolor{black}{$0.13$} & \textcolor{black}{$0.13$}\\
$0.50$ & $10$ & $-14.59$ & $-14.59\pm0.12$ & $0.12$ & $0.01$ & \textcolor{black}{$0.02$} & \textcolor{black}{$0.12$} & \textcolor{black}{$0.50$} & \textcolor{black}{$0.48$}\\
$0.50$ & $30$ & $-60.13$ & $-60.12\pm0.25$ & $0.24$ & $0.02$ & \textcolor{black}{$0.03$} & \textcolor{black}{$0.29$} & \textcolor{black}{$0.56$} & \textcolor{black}{$0.55$}\\
$0.50$ & $50$ & $-112.42$ & $-112.33\pm0.34$ & $0.33$ & $0.03$ & \textcolor{black}{$0.27$} & \textcolor{black}{$2.65$} & \textcolor{black}{$0.40$} & \textcolor{black}{$0.58$}\\
$1$ & $2$ & $-1.75$ & $-1.75\pm0.05$ & $0.04$ & $0.00$ & \textcolor{black}{$-0.02$} & \textcolor{black}{$-0.30$} & \textcolor{black}{$0.01$} & \textcolor{black}{$0.01$}\\
$1$ & $10$ & $-14.59$ & $-14.59\pm0.12$ & $0.12$ & $0.01$ & \textcolor{black}{$0.02$} & \textcolor{black}{$0.19$} & \textcolor{black}{$0.49$} & \textcolor{black}{$0.61$}\\
$1$ & $30$ & $-60.13$ & $-60.46\pm0.25$ & $0.23$ & $0.02$ & \textcolor{black}{$-1.36$} & \textcolor{red}{$-14.57$} & \textcolor{black}{$0.48$} & \textcolor{black}{$0.53$}\\
$1$ & $50$ & $-112.42$ & $-113.52\pm0.34$ & $0.32$ & $0.03$ & \textcolor{red}{$-3.26$} & \textcolor{red}{$-34.47$} & \textcolor{black}{$0.50$} & \textcolor{black}{$0.51$}\\
$2$ & $2$ & $-1.75$ & $-1.74\pm0.05$ & $0.06$ & $0.01$ & \textcolor{black}{$0.06$} & \textcolor{black}{$0.43$} & \textcolor{red}{$6.1\cdot10^{-6}$} & \textcolor{red}{$2.1\cdot10^{-5}$}\\
$10$ & $10$ & $-14.59$ & $-14.05\pm0.12$ & $0.36$ & $0.04$ & \textcolor{red}{$4.42$} & \textcolor{red}{$15.01$} & \textcolor{black}{$0.09$} & \textcolor{black}{$0.09$}\\
$30$ & $30$ & $-60.13$ & $-38.78\pm0.21$ & $1.34$ & $0.13$ & \textcolor{red}{$103.26$} & \textcolor{red}{$159.47$} & \textcolor{red}{$3.5\cdot10^{-5}$} & \textcolor{red}{$5.2\cdot10^{-3}$}\\
$50$ & $50$ & $-112.42$ & $-64.20\pm0.63$ & $5.17$ & $0.52$ & \textcolor{red}{$103.63$} & \textcolor{red}{$93.31$} & \textcolor{red}{$5.2\cdot10^{-12}$} & \textcolor{red}{$3.8\cdot10^{-7}$}\\
\hline
\hyperref[sec:gaussian-log-gamma]{Mixture}\\
\hline
$0.50$ & $2$ & $-8.19$ & $-8.17\pm0.06$ & $0.06$ & $0.01$ & \textcolor{black}{$0.35$} & \textcolor{red}{$3.33$} & \textcolor{black}{$0.45$} & \textcolor{black}{$0.56$}\\
$0.50$ & $10$ & $-40.94$ & $-40.87\pm0.16$ & $0.15$ & $0.02$ & \textcolor{black}{$0.49$} & \textcolor{red}{$5.12$} & \textcolor{black}{$0.43$} & \textcolor{black}{$0.52$}\\
$0.50$ & $20$ & $-81.89$ & $-81.75\pm0.23$ & $0.24$ & $0.02$ & \textcolor{black}{$0.61$} & \textcolor{red}{$5.70$} & \textcolor{black}{$0.51$} & \textcolor{black}{$0.51$}\\
$1$ & $2$ & $-8.19$ & $-8.16\pm0.06$ & $0.06$ & $0.01$ & \textcolor{black}{$0.39$} & \textcolor{red}{$3.78$} & \textcolor{black}{$0.44$} & \textcolor{black}{$0.49$}\\
$1$ & $10$ & $-40.94$ & $-40.85\pm0.16$ & $0.16$ & $0.02$ & \textcolor{black}{$0.59$} & \textcolor{red}{$5.71$} & \textcolor{black}{$0.48$} & \textcolor{black}{$0.45$}\\
$1$ & $20$ & $-81.89$ & $-81.72\pm0.22$ & $0.27$ & $0.03$ & \textcolor{black}{$0.73$} & \textcolor{red}{$6.18$} & \textcolor{black}{$0.52$} & \textcolor{black}{$0.59$}\\
$2$ & $2$ & $-8.19$ & $-8.18\pm0.06$ & $0.09$ & $0.01$ & \textcolor{black}{$0.09$} & \textcolor{black}{$0.58$} & \textcolor{black}{$0.52$} & \textcolor{black}{$0.49$}\\
$10$ & $10$ & $-40.94$ & $-40.90\pm0.18$ & $1.31$ & $0.13$ & \textcolor{black}{$0.77$} & \textcolor{black}{$0.31$} & \textcolor{red}{$7.2\cdot10^{-3}$} & \textcolor{black}{$0.27$}\\
$20$ & $20$ & $-81.89$ & $-88.28\pm0.64$ & $5.80$ & $0.58$ & \textcolor{red}{$-11.25$} & \textcolor{red}{$-11.03$} & \textcolor{red}{$4.5\cdot10^{-15}$} & \textcolor{black}{$0.01$}\\
\end{tabular}
}
\caption{\label{tab:PC_summary} Summary of results of our insertion index cross-check for \PC. See \cref{tab:MN_summary} for further details. In this table we show $d / \nr$, which may be thought of as a ``\PC efficiency'' analogue of the \MN efficiency $\efr$.}
\end{table*}
\endgroup

\subsection{Toy functions}

We now present detailed numerical examples of our cross-check using NS runs on toy functions using \MNVersion~\citep{Feroz:2007kg,Feroz:2008xx,Feroz:2013hea} and \PCVersion~\citep{Handley:2015fda,Handley:2015xxx}. We chose toy functions with known analytic evidences or precisely known numerical estimates of the evidence to demonstrate that biased results from NS are detectable with our approach. The toy functions are described in \cref{app:toy_problems}.

We performed \nrepeatSetting \MN and \PC runs on each toy function to study the statistical properties of their outputs. We used $\nlive = \nliveSetting$ and $\stoppingtol = \tolSetting$ throughout. To generate biased NS runs, we used inappropriate settings, e.g., $\efr > 1$ in \MN or few repeats $\nr<d$ in slice sampling in \PC, and difficult toy functions with $d \ge 30$. We post-processed the results using \anesthetic~\citep{anesthetic}. 

We summarise our results by the average \logZ and error estimate $\Delta \logZ$, and by the median \pvalue from all the insertion indexes and the median running \pvalue. We furthermore report the standard error on the mean, SEM \logZ, and the standard deviation, $\sigma_{\logZ}$. We use the error estimates to compute the average inaccuracy and bias,
\begin{align}
\text{inaccuracy} ={}& \frac{\logZ - \text{analytic}}{\Delta\logZ},\\ 
\text{bias} ={}& \frac{\logZ - \text{analytic}}{\text{SEM} \logZ}.
\end{align}
The inaccuracy shows whether the uncertainty reported by a code from single runs was reasonable.

We present our numerical results using \MN and \PC in tables~\ref{tab:MN_summary} and \ref{tab:PC_summary}, respectively. First, for the Gaussian function, the \MN estimates of \logZ were significantly biased for $d = 30$ and $50$ for all \efr settings, and for $d = 2$ and $10$ for $\efr = 10$. Our cross-check was successful, as the \pvalues corresponding to the biased results were tiny. 

For the Rosenbrock function, our cross-check detected a problem with \MN runs with $d=2$ and $\efr = 10$, even though the \MN evidence estimate was not biased. It did not detect a problem with $\efr=1$, even though the \logZ estimate was biased.  This was, however, the only problem for which this occurred for \MN.

For the shells function, the \MN estimates of \logZ were biased for many combinations of $d$ and \efr. The biased results were all identified by our cross-check with tiny \pvalues. Indeed, when $d=50$, even with $\efr=0.1$, we saw a bias of about $115$ and a median rolling \pvalue of about $10^{-23}$.

Lastly, the $d = 20$ mixture functions are particularly important, as \MN was known to produce biased results even with $\efr = 0.1$. Using all the insertion indexes, we find $\pvalue \approx 10^{-6}$ for this function, i.e., our cross-check successfully detects these failures.

In the analogous results for \PC in \cref{tab:PC_summary} we see fewer significantly biased estimates throughout, and only three biased results when using the recommended $\nr = 2d$ setting, which all occurred in the mixture function. We note, though, that the error estimates from \PC were reasonable even in these cases. The most extremely biased results were detected by our cross-check in the Gaussian, shells and mixture functions. 

Our cross-check detected faults in the $d=2$ shells function for $\nr = 1$, despite no evidence of bias in \PC results. Perhaps this should not be surprising, as \citep{2018arXiv180503924S,2020MNRAS.tmp..280S} suggest that independent samples from the constrained prior are not strictly necessary for correct evidence estimates.
The \pvalues, however, increased monotonically as \nr was increased, as expected. Lastly, we note that in many more cases than for \MN biases were not detected by our cross-check; this may be because the biases are smaller than they were for \MN.

In summary, for both \MN and \PC, we find that our cross-check can detect problematic NS runs in a variety of functions, settings and dimensions, although there is room for refinement. The problem detected by our cross-check usually leads to a faulty estimate of the evidence, though in a few cases the evidence estimate remains reasonable despite the apparent failure to sample correctly from the constrained prior.

\subsection{Cosmological model selection}

In \citep{Handley:2019tkm}, Handley considered the Bayesian evidence for a spatially closed Universe. Bayesian evidences from combinations of four datasets were computed using \PC for a spatially flat Universe and a curved Universe. The resulting Bayes factors showed that a closed Universe was favoured by odds of about $50/1$ for a particular set of data. There were $22$ NS computations in total. The \PC results are publicly archived at \citep{will_handley_2019_3371152}. We ran our cross-check on each of the $22$ NS runs in the archived data, finding \pvalues in the range $4\%$ to $98\%$. The results do not suggest problems with the NS runs. The $\pvalue$ of $4\%$  is not particularly alarming, especially considering that  we conducted $22$ tests. The full results are shown in \cref{tab:cosmology}.

\begingroup
\begin{table*}
\begin{tabular}{lcccc}
& \multicolumn{2}{c}{Flat} & \multicolumn{2}{c}{Curved}\\
\cline{2-3}\cline{4-5}
Data & \pvalue & Rolling \pvalue& \pvalue & Rolling \pvalue\\
\hline
BAO & 0.89 & 0.82 & 0.07 & 0.05\\
lensing+BAO & 0.72 & 0.54 & 0.19 & 0.43\\
lensing & 0.26 & 0.14 & 0.04 & 0.64\\
lensing+S$H_0$ES & 0.08 & 0.08 & 0.78 & 0.04\\
Planck+BAO & 0.39 & 0.56 & 0.14 & 0.43\\
Planck+lensing+BAO & 0.68 & 0.69 & 0.70 & 0.27\\
Planck+lensing & 0.94 & 0.49 & 0.89 & 0.72\\
Planck+lensing+S$H_0$ES & 0.92 & 0.92 & 0.33 & 0.82\\
Planck & 0.81 & 0.69 & 0.84 & 0.88\\
Planck+S$H_0$ES & 0.20 & 0.48 & 0.92 & 0.97\\
S$H_0$ES & 0.59 & 0.59 & 0.98 & 0.98\\
\end{tabular}
\caption{\label{tab:cosmology}Insertion index cross-check applied to NS results from cosmological model selection in \citep{Handley:2019tkm}. We show \pvalues and rolling \pvalues for the NS evidence calculations for flat and curved Universe models with 11 datasets. See \citep{Handley:2019tkm} for further description of the datasets and models.}
\end{table*}
\endgroup

\subsection{Plateaus}

Let us consider the  one-dimensional function in example 2 from \citep{2020arXiv200508602S}. The likelihood function is defined piece-wise to be a Gaussian at the center and zero in the tails;
\begin{equation}
\like(x) \propto
\begin{cases} 
      e^{-\frac{(x - \mu)^2}{2\sigma^2}} & \left(\frac{x - \mu}{\sigma}\right)^2 \leq 1\\
      0 & \text{elsewhere}. 
   \end{cases}
\end{equation}
for $\mu = \tfrac12$ and $\sigma = 1$. The prior is uniform from $-3$ to $3$. We confirm that the NS algorithm produces biased estimates of the evidence in this function. However, since the likelihood is zero in $5/6$ of the prior, approximately $5/6$ of the initial live points have a likelihood of zero and share the same insertion index from \cref{eq:insertion_index}. This results in a tiny $\pvalue \simeq 0$ in our test.

\subsection{Perfect NS}

Lastly, we simulated perfect NS runs that correctly sample from the constrained prior. We simulated them by directly sampling compression factors from uniform distributions and never computing any likelihoods. Of course, with no likelihood we cannot compute an evidence, but we can simulate insertion indexes. We performed \nrepeatsPerfectSetting runs of perfect NS with \niterPerfectSetting iterations and computed the \pvalue via our KS test. 

\begin{figure}[t]
\centering
\includegraphics[width=0.9\linewidth]{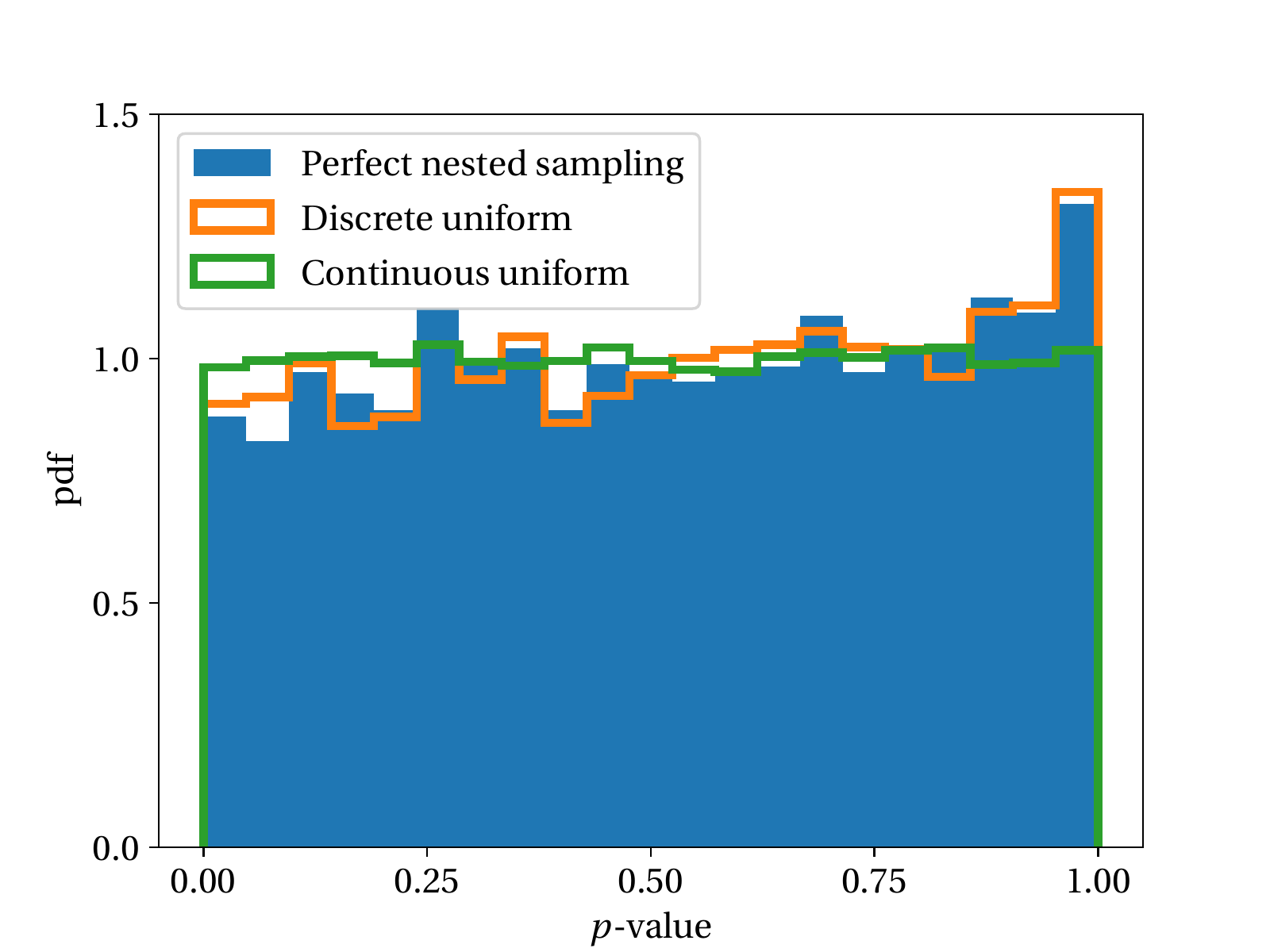}
\caption{Histogram of \pvalue{}s from tests uniformity of insertion indexes from perfect NS (blue), samples from a discrete uniform distribution (orange) and samples from a continuous uniform distribution (green).}\label{fig:p_value_histogram}
\end{figure}

We furthermore computed \nrepeatsMCSetting \pvalue{}s from a KS test on \niterPerfectSetting samples drawn from a continuous uniform distribution and on \niterPerfectSetting samples drawn from a discrete uniform distribution with \nliveSetting bins.  We histogram all \pvalue{}s in \cref{fig:p_value_histogram}. Of course, the KS \pvalue{}s should be uniformly distributed in the continuous case and it appears that it is (green). The impact of discretization on the KS test is visible (orange) but small with \nliveSetting live points. The further impact of correlations amongst the samples in perfect NS (blue) isn't obvious. A two-sample Kolmogorov-Smirnoff test similar to the one-sample test in \cref{app:ks} uncovered a slight difference between \pvalues from perfect NS and from a discrete uniform ($\text{\pvalue} = 0.03$). This suggests that although the correlations and discretization impact the KS test, the effect is small. 

\section{Future use of insertion indexes}\label{sec:debiasing}

For the purposes of evidence estimation, a nested sampling run is fully encoded by recording the {\em birth contour\/} and {\em death contour\/} of each point~\citep{higson2018sampling}. For the purposes of estimating volume in a statistical way, we generally discard the likelihood information, focussing on the ordering of the contours. This makes sense, as barring the stopping criterion in \cref{eqn:stop}, the underlying nested sampling algorithm is athermal and insensitive to monotonic transformations of the likelihood.

Traditional nested sampling uses the fact that
\begin{equation}
    P(X_j|X_{j-1}, n_\mathrm{live}) = \frac{n_j}{X_{j-1}}\left( \frac{X_j}{X_{j-1}} \right)^{n_j-1} [0<X_j<X_{j-1}].
    \label{eqn:prob}
\end{equation}
In the above, one has essentially marginalised out dependency on everything other than $X_{j-1}$, and compressed the birth-death contour information into a vector encoding the number of live points at each iteration $n_i$. One can then use this recursively (alongside the fact that $X_0=1$) to perform inference on $P(X)$ and therefore the evidence via \cref{eq:t,eq:Z_sum}.

The critical question therefore is whether this ``Skilling compression'' from birth-death contours to numbers of live points is lossless or lossy for the purposes of volume estimation (note that it is generically lossy, as it's impossible to go in the reverse direction). The results presented in this paper are suggestive that it is losing some useful information, as insertion indexes do provide further information in the context of a cross check (and are in fact a lossless compression of the birth and death contours). One possibility is that the Skilling compression is lossless in the context of perfect nested sampling, but if a run is biased then you may be able to use insertion indexes to partially correct a biased run. This is the subject of ongoing work by the authors.

\section{Conclusions}\label{sec:conclusions}

We identified a previously unknown property of the NS algorithm: the insertion indexes of new live points into the existing live points should be uniformly distributed. This observation enabled us to invent a cross-check of single NS runs. The cross-check can detect when an NS run fails to sample new live points from the constrained prior, which is the most challenging aspect of an efficient implementation of NS, and functions with plateaus in the likelihood function recently identified in \citep{2020arXiv200508602S}, both of which can lead to unreliable estimates of the evidence and posterior, 

We applied our cross-check to NS runs on several toy functions with known analytic results in $2$ -- $50$ dimensions with \MN and \PC, which sample from the constrained prior using ellipsoidal rejection sampling and slice sampling, respectively. Our numerical results are some of the most detailed checks of \MN and \PC. We found that our cross-check could detect problematic runs for both codes. Since the idea is relatively simple, we suggest that a cross-check of this kind should become a mandatory test of any NS run. The exact form of the cross-check, however, could be refined. We chose a KS test using all the iterations or the most significant $\nlive$ iterations; both choices could be improved. As an example of a realistic application, we furthermore applied our cross-check to results from $22$ NS runs performed in the context of cosmological model selection.

Lastly, we speculated that the information contained in the insertion indexes could be used to debias single NS runs or lead to an improved formula for the evidence summation. We outlined a few difficulties and hope our observations lead to further developments.

Future work will involve extending the method to work in the context of a variable number of live points, as well as exploring the larger possibilities of using order statistics to improve NS accuracy and potentially debias runs.

\section*{Acknowledgments}
The authors would like to thank Gregory Martinez for valuable discussions. We thank the organisers of the GAMBIT XI workshop where some of this work was planned and completed. AF was supported by an NSFC Research Fund for International Young Scientists grant 11950410509.
WH was supported by a George Southgate visiting fellowship grant from the University of Adelaide, and STFC IPS grant number G102229.

\section*{Data availability}
The raw data from our \nrepeatSetting runs for \cref{tab:MN_summary,tab:PC_summary} are publicly available at \citep{fowlie_andrew_2020_3958749}. All other data will be shared on reasonable request to the corresponding author.

\bibliographystyle{mnras}
\bibliography{references}

\appendix
\section{Proof of \cref{EQ:UNIFORM}}\label{app:proof}  

In NS we have $n = \nlive - 1$ remaining samples after the worst live point was removed. Their associated volumes were drawn from a (continuous) uniform distribution, $X_i \sim \uniform(0, 1)$. If we draw another sample, the distribution of its insertion index, $i$, relative to the other samples depends on the probability contained in the uniform distribution between the ordered samples. In fact, the probability for each insertion index $i = 0, 1, \ldots, n$ is
\begin{align}
P_i ={}& \int (X_{i+1} - X_i) \, p(X_{i+1}, X_i) \,\intd X_{i+1} \,\intd X_i\\
    ={}& \int X_{i+1} \,  p(X_{i+1}) \,\intd X_{i+1} - \int X_{i} \, p(X_i) \,\intd X_{i}\\
    ={}& \langle X_{i+1} \rangle - \langle X_{i} \rangle,
\end{align}
where we completed two trivial integrals and wrote the terms as expectations. To compute the expectations, note that
\begin{equation}
p(X_i) = \frac{n!}{(i - 1)! (n - i)!} (1 - X_i)^{n - i} X_i^{i - 1},
\end{equation}
since we need $n - i$  samples above $X_i$, $i - 1$ samples below $X_i$ and one sample at $X_i$. The first factor is combinatoric; the second accounts for the $n - i$ samples that must lie above $X_i$; and the third accounts for the $i - 1$ samples that must lie below $X_i$. The factor for a final sample at $X_i$ is just one. By integration, we quickly find $\langle X_i \rangle = i / (n + 1)$, and thus $P_i = 1 / (n + 1)$. That is, the insertion indexes follow a discrete uniform distribution.

Note that this didn't depend especially on the fact that the distribution of the samples was uniform. If the samples had followed a different distribution, we can transform $X_i \to Y_i = F(X_i)$ where $F$ is the cumulative distribution function, such that $Y_i \sim \uniform(0, 1)$, the proof goes through just the same.

\section{Kolmogorov-Smirnov test}\label{app:ks}

We use a one-sample Kolmogorov-Smirnov (KS) test~\citep{kolmogorov1933sulla,smirnov1948} to compare our set of $\niter$ insertion indexes with a (discrete) uniform distribution. First, we compute the KS test-statistic by comparing the empirical cumulative distribution function, $F_\text{data}$, to that from a discrete uniform distribution, $F_U$,
\begin{equation}\label{eq:Dn}
D_n =  \sup_x \left|F_\text{data}(x) - F_U(x)\right|.
\end{equation}
This provides a notion of distance between the observed indexes and a uniform distribution. 
In the continuous case, the null-distribution of this test-statistic does not depend on the reference distribution. We convert the test-statistic into a \pvalue using an asymptotic approximation of the Kolmogorov distribution~\citep{JSSv008i18} implemented in \code{scipy}~\citep{2020SciPy-NMeth},
\begin{equation}
\text{\pvalue} = P\left(D_n^{\phantom{\star}} \ge D_n^\star \,\mid\, H_0\right),
\end{equation}
where $D_n^\star$ is the observed statistic. This assumes that we are testing samples from a continuous distribution. In our discrete case, the \pvalues from the Kolmogorov distribution are known to be conservative~\citep{RJ-2011-016}.

\section{Toy functions}\label{app:toy_problems}

\subsection{Gaussian}\label{sec:gaussian}
Our first example is a multi-dimensional Gaussian likelihood,
\begin{equation}
\like(\params) = \frac{1}{\sqrt{(2\pi)^n \det \Sigma}} e^{-\tfrac12 (\params - \bm{\mu})^T \Sigma^{-1} (\params - \bm{\mu})},
\end{equation}
with covariance matrix $\Sigma$ and mean $\bm{\mu}$. We pick a uniform prior from $0$ to $1$ for each dimension. The analytic evidence is always $\logZ = 0$ since the likelihood is a pdf in $\params$, modulo small errors as the infinite domain is truncated by the prior. We pick $\mu = 0.5$ and a diagonal covariance matrix with $\sigma = 0.001$ for each dimension.

\subsection{Rosenbrock}\label{sec:rosenbrock}
This is a two-dimensional function exhibiting a pronounced curved degeneracy~\cite{10.1093/comjnl/3.3.175}. The likelihood function is
\begin{equation}
-\ln\like(x, y) = (1 - x)^2  + 100 (y - x^2)^2.
\end{equation}
We consider uniform priors from $-5$ to $5$ for each parameter. The evidence can be found semi-analytically from a one-dimensional integral,
\begin{equation}
\Z = \frac{\sqrt{\pi} }{2000} \, \int_{-5}^5 \left[\text{erf}\left(10 (5 -x^2)\right) + \text{erf}\left(10 (5 +x^2)\right)\right] e^{-(1 - x)^2} \, \intd x
\end{equation}
to be $\logZ = -5.804$. The analytic approximation, which approximates the $y$ domain of integration by the whole real line, leads to
\begin{equation}
\Z \approx \frac{\pi}{20000} \left[\text{erf}(6) - \text{erf}(4)\right],
\end{equation}
and thus $\logZ = -5.763$.

\subsection{Gaussian shells}\label{sec:shells}

The multidimensional likelihood is
\begin{equation}
\like(\params) = \text{shell}(\params; \bm{c}, r, w) + \text{shell}(\params; -\bm{c}, r, w)
\end{equation}
where the shell function is a Gaussian favouring a radial distance $r$ from the point $\bm{c}$,
\begin{equation}
\text{shell}(\params; \bm{c}, r, w) = \frac{1}{\sqrt{2\pi} w} e^{-(|\params - \bm{c}| - r)^2 / (2 w^2)}.
\end{equation}
Thus, the highest likelihood region forms a shell of characteristic width $w$ at the surface of a $d$-sphere of radius $r$. Our likelihood contains two such shells, one at $\bm{c}$ and one at $\bm{-c}$. As usual, we take $w = 0.1$, $r = 2$ and $\bm{c} = (3.5, 0, \ldots, 0)$.

With uniform priors between $-6$ and $6$, the analytic evidence is approximately,
\begin{equation}
\Z = 2  \langle |x|^{d-1}\rangle S_d / 12^d
\end{equation}
where $S_d$ is the surface area of an $d$-sphere and $\langle |x|^{d-1}\rangle$ is the $(d-1)$-th non-central moment of a Gaussian, $\normal(r, w^2)$, and we ignore the truncation of the domain by the finite-sized hypercube.

\subsection{Gaussian-Log-Gamma mixture}\label{sec:gaussian-log-gamma}
This toy function was found in~\cite{2013arXiv1304.7808B,Feroz:2013hea,2014arXiv1407.5459B} to be problematic in \MN without importance sampling. It is defined in even numbers of dimensions.  The likelihood is a product of $d$ factors,
\begin{equation}
\like(\params) = \prod_{i=1}^d \like_i(\theta_i),
\end{equation}
where the factors are
\begin{align}
\like_i(\theta) =
\begin{cases} 
      \tfrac12 \loggamma(\theta \rvert 10, 1, 1) + \tfrac12 \loggamma(\theta \rvert {-10}, 1, 1) & i = 1\\
      \tfrac12 \normal(\theta \rvert 10, 1) + \tfrac12 \normal(\theta \rvert {-10}, 1) & i = 2\\
      \loggamma(\theta \rvert 10, 1, 1) & 3 \le i \le \frac{d+2} {2}\\
      \normal(\theta \rvert 10, 1) & \frac{d+2}{2}  < i \le d\\
   \end{cases}
\end{align}
where e.g., $\loggamma(\theta \rvert 10, 1, 1)$ denotes a one-dimensional log-Gamma density for $\theta$ with mean $10$ and shape parameters $1$ and $1$.
There are four identical modes at $\theta_1 = \pm 10$, $\theta_2 = \pm10$ and $\theta_{i>2} = 10$.

The prior is uniform in each parameter from $-30$ to $30$. Since the likelihood is a pdf in $\params$, the analytic $\logZ$ is governed by the prior normalization factor, $\logZ = \log(1/60^d) \approx -4.1 d$, modulo small truncation errors introduced by the prior.

\bsp	
\label{lastpage}
\end{document}